\documentstyle[aps,epsf,preprint]{revtex}
\def\be{\begin{equation}}
\def\ee{\end{equation}}
\def\bea{\begin{eqnarray}}
\def\eea{\end{eqnarray}}
\def\beq{\begin{equation}}
\def\eeq{\end{equation}}
\begin{document}
\draft
\preprint{}
\title{Squeezed Fermions at Relativistic Heavy Ion Colliders \\[2ex]}
\author{P. K. Panda$^1$, T. Cs\"org\H o$^{1,2,3,4}$, 
Y. Hama$^3$, G. Krein$^1$ and Sandra S. Padula$^1$\\[2ex]}
\address{$^1$ Instituto de F\'{\i}sica Te\'orica, 
Universidade Estadual Paulista, \\
Rua Pamplona 145, 01405-900 S\~ao Paulo - SP, Brazil,\\
$^2$ MTA KFKI RMKI, H - 1525 Budapest 114, POB 49, Hungary,\\
$^3$ Instituto de F\'{\i}sica, Universidade de S\~ao Paulo, \\
C.P. 66318 - CEP 05315-970 S\~ao Paulo,  Brazil\\
$^4$ Department of Physics, Columbia University, 538 W 120th St,\\
New York, NY 10027 }
\maketitle
\begin{abstract}
Large back-to-back correlations of observable fermion -- anti-fermion pairs 
are predicted to appear, if the mass of the fermions is modified in a 
thermalized medium. The back-to-back correlations of protons and anti-protons 
are experimentally observable in ultra-relativistic heavy ion collisions, 
similarly to the Andreev reflection of electrons off the boundary of a 
superconductor. While quantum statistics suppresses the probability of 
observing pairs of fermions with nearby momenta, the fermionic
back-to-back correlations are positive and of similar strength to bosonic 
back-to-back correlations.
\end{abstract}

\vspace{1.0cm}

\section{Introduction} In high energy heavy ion collisions, hot and dense 
hadronic matter is expected to be created in conditions similar to the ones 
in the early Universe about a few $\mu$sec after the Big Bang. Under these 
conditions, strongly interacting particles may propagate with a mass that 
differs from the mass in the asymptotic vacuum. Recently, it has been  
discovered that in-medium mass-modifications induce large back-to-back 
correlations (BBC) among pairs of asymptotic, observable 
bosons~\cite{ac,acg}. However, it was not known before if the bosonic 
BBC would have an analogous effect in the fermion sector. In this Letter we 
show that fermion - anti-fermion pairs also exhibit BBC, if they propagate 
with a modified  mass in the medium. The physical picture is that of a
system in thermodynamical equilibrium which freezes-out suddenly. The 
masses and other properties of the hadrons in the system are modified due to 
interactions and a quasi-particle description is assumed. At freeze-out the 
quasi-particles are converted into the free particles that will be detected. 
Particle--anti-particle pairs detected with momenta back-to-back give rise 
to a quantum-mechanical correlation due to a nonzero overlap between the 
in-medium states and the free states. The quantum correlation can be 
calculated with a Bogoliubov-Valatin transformation between a quasi-particle 
basis and a free-particle basis. 

The bosonic BBC (bBBC) have a quantum optical analogy, namely the correlations
in thermalized ensembles of two-mode squeezed states. It turns out that 
fermion BBC (fBBC) 
also have an analogy, which is the Andreev reflection, well-known in solid 
state physics. It refers to the scattering of electrons off the boundary of 
a superconductor - normal conductor junction. The reflected electrons are used
to study the properties of the superconductors~\cite{andreev}. Our results 
for fBBC generalize Andreev's reflection to the case when superconductivity 
is suddenly switched off in the whole volume of the material, so that a 
junction prevails not at a given position for a long time but in the whole 
medium at a given instant. 

Quantum statistics enhances the probability of observing pairs of bosons 
in similar momentum states, while it suppresses the probability of observing
pairs of fermions with nearby momenta. In spite of such Fermi-Dirac 
suppression factors that appear in the evaluation of the correlation function,
we find that the dominant term of the fBBC,  due to the Bogoliubov-Valatin 
transformation, is positive and similar in strength to the bosonic BBC of 
Refs.~\cite{ac,acg}. The fBBC could be observed 
experimentally in $^{197}$Au + $^{197}$Au collisions at the Relativistic 
Heavy Ion Collider (RHIC), which started to take data at the Brookhaven 
National Laboratory at $\sqrt{s} = 56$ and $130$ AGeV in 2000 and which will 
reach its full designed energy of $ \sqrt{s} = 200$ AGeV in  the near future.

The outline of the paper is as follows. In section II we consider
the in-medium mass modification state at finite temperature. The spectra
and the correlations for mass-shifted fermions are described in section 
III and section IV contains concluding remarks.

\section{Basic assumptions} We assume the validity of the concepts of local 
thermalization, hydrodynamics, and a short duration of particle emission. 
These concepts are in agreement with the observable single-particle spectra 
and two-particle correlations  of pions, kaons and protons~\cite{bl-qm99}.
We also assume the validity of the effective Hamiltonian 
\be
H=H_0+H_I ,
\label{H} \label{e:ham}
\ee
where 
\be
{H}_0 =  \int d{\bf x}\, :\bar\psi({\bf x})(-i {\bbox \gamma} \cdot 
{\bbox \nabla} + M) \psi({\bf x}):
\label{H0}
\ee
is the Hamiltonian in vacuum, and the interaction Hamiltonian $H_I$ describes 
the medium modifications and will be discussed shortly below. In 
Eq. (\ref{H0}), $M$ is the value of the proton (or in general, baryon) mass 
in free space, $\psi$ and $\bar \psi$ are the fermion field operators which 
satisfy the usual equal-time anti-commutation relations. 

We are interested in the single-particle and two-particle invariant momentum 
distributions for fermions and anti-fermions:
\begin{eqnarray}
&& N_1({\bf k}_1) = \omega_{{\bf k}_1} \langle
a^\dagger_{{\bf k}_1} a_{{\bf k}_1}\rangle , ~~~~~~~
{\tilde N}_1({\bf k}_1)  = \omega_{{\bf k}_1} \langle
 {\tilde a}^\dagger_{{\bf k}_1} {\tilde a}_{{\bf k}_1}\rangle , \label{N1s} \\
&& N_2({\bf k}_1,{\bf k}_2) = 
\omega_{{\bf k}_1} \omega_{{\bf k}_2} \langle a^\dagger_{{\bf k}_1} 
{\tilde a}^\dagger_{{\bf k}_2} {\tilde a}_{{\bf k}_2} a_{{\bf k}_1}\rangle 
\label{N2}.
\end{eqnarray}
Here, $\langle \hat O \rangle$ denotes the expectation value of operator 
$\hat O$ in the thermalized medium, and the $a^\dagger$, $a$, 
$\tilde a^\dagger$, $\tilde a$ are creation and annihilation operators of 
free baryons and anti-baryons of mass $M$ and energy $\omega_{\bf k} = 
\sqrt{M^2 + |{\bf k}|^2 }$. These creation and annihilation operators 
are defined through the expansion of the baryon field operator as 
\be
\psi({\bf x})= \frac{1}{\sqrt{V}} \sum_{\lambda,\lambda^\prime,{\bf k}} 
\left( u_{\lambda,{\bf k}} a_{\lambda,{\bf k}} + v_{\lambda^\prime,-{\bf k}} 
{\tilde a}_{\lambda^\prime,-{\bf k}}^\dagger \right) e^{i{\bf k}\cdot{\bf x}} ,
\label{fe}
\ee
where $V$ is the volume of the system, $u_{\lambda,{\bf k}}$ and 
$v_{\lambda^\prime,-{\bf k}}$ are Dirac spinors and the summation 
extends over momenta ${\bf k}$ and spin projections $\lambda,\lambda^\prime
=1/2,-1/2$. The creation and annihilation operators satisfy canonical 
anti-commutation relations. The particle--anti-particle correlation function 
is defined as 
\begin{equation}
C_2({\bf k}_1,{\bf k}_2) = \frac{N_2({\bf k}_1,{\bf k}_2)}
{N_1({\bf k}_1) {\tilde N}_1({\bf k}_2)} 
\;. \label{e:hbt}
\end{equation}

As in the bosonic case~\cite{ac,acg}, the expectation value of the four baryon
operators in Eq.~(\ref{N2}) can be calculated using Wick's theorem generalized
for a locally equilibrated system, with the result
\begin{eqnarray}
\langle a^\dagger_{{\bf k}_1} {\tilde a}^\dagger_{{\bf k}_2} 
{\tilde a}_{{\bf k}_2} a_{{\bf k}_1}\rangle=
\langle a^\dagger_{{\bf k}_1} a_{{\bf k}_1}\rangle
\langle  {\tilde a}^\dagger_{{\bf k}_2} {\tilde a}_{{\bf k}_2} \rangle 
- \langle a^\dagger_{{\bf k}_1} {\tilde a}_{{\bf k}_2}\rangle
\langle {\tilde a}^\dagger_{{\bf k}_2} a_{{\bf k}_1}\rangle 
+ { \langle a^\dagger_{{\bf k}_1}{\tilde a}^\dagger_{{\bf k}_2}\rangle
 \langle  {\tilde a}_{{\bf k}_2} a_{{\bf k}_1} \rangle}.
\label{rand}
\end{eqnarray}
The minus sign in the above equation is due to Fermi statistics. The 
expectation values involving $a^{\dagger}_{{\bf k}_1} a_{{\bf k}_2}$ and  
$\tilde a^{\dagger}_{{\bf k}_1} \tilde a_{{\bf k}_2}$ give rise to the chaotic
amplitudes, while $a_{{\bf k}_1} \tilde a_{{\bf k}_2}$ gives rise to the
``squeezed'' amplitude, defined as
\begin{eqnarray}
G_c({\bf k}_1,{\bf k}_2) & = & \sqrt{\omega_{{\bf k}_1} \omega_{{\bf k}_2}}
\langle a^{\dagger}_{{\bf k}_1} a_{{\bf k}_2}\rangle~,~~~~~~
{\tilde G}_c({\bf k}_1,{\bf k}_2)  =  {\sqrt{\omega_{{\bf k}_1} 
\omega_{{\bf k}_2}}} \langle {\tilde a}^{\dagger}_{{\bf k}_1} 
{\tilde a}_{{\bf k}_2}\rangle,\label{gc}\\
G_s({\bf k}_1,{\bf k}_2) & =  &  \sqrt{\omega_{{\bf k}_1} \omega_{{\bf k}_2} }
\langle a_{{\bf k}_1} {\tilde a}_{{\bf k}_2} \rangle~~.
\label{gs}
\end{eqnarray}
The chaotic amplitudes are non-vanishing for identical type of fermions and 
anti-fermions, while the squeezed amplitude is non-vanishing only for 
particle-anti-particle pairs. We will show in the following that, as in the 
case of bosons~\cite{ac,acg}, the squeezed $G_s({\bf k}_1,{\bf k}_2)$ can be 
considerably large if the in-medium masses of the baryons are different from 
their free-space values. 

In order to evaluate the thermal averages above we model the system as a 
globally thermalized gas of quasi-particles (quasi-baryons). The density matrix
for such a system is given not in terms of the free-space creation and 
annihilation operators $a^{\dagger}$, $\tilde a$, $\cdots$ , but by a new set 
that we denote by $b^{\dagger}$, $\tilde b$, $\cdots$. They are of course 
different because of medium effects, described by the interacting Hamiltonian 
in Eq.~(\ref{e:ham}). While it is the $a$-quanta that are observed as 
asymptotic states, it is the $b$ quanta that are thermalized in the medium. 

In a quasi-particle description of the system, the medium effects are taken 
into account through a self-energy function. For a spin-1/2  particle under 
the influence of mean fields in a many-body system, one can write its 
self-energy function as
\be
\Sigma = \Sigma^s + \gamma^0 \Sigma^0 + \gamma^i \Sigma^i .
\ee
This function can be determined from a detailed calculation based on, for 
example, the Walecka or the Zim\'anyi - Moszkowski models~\cite{wzm,franc}. 
Calculations in the context of such models show that $\Sigma^0$ is weakly 
momentum dependent and $\Sigma^i$ is very small (see for example 
Ref.~\cite{franc} for detailed numerical results). It is well known that a 
tensor contribution may also occur, however, in applications for nuclear 
matter and finite nuclei this additional component is negligible, and will 
not be considered here. Therefore, we assume $\Sigma^0$ independent of 
momentum, and neglect $\Sigma^i$. We denote the scalar component of the 
self-energy as $\Sigma^s ({\bf k})  = \Delta M({\bf k})$. Now, for a locally 
thermalized system, the role of $\Sigma^0$ is to shift the chemical 
potential, i.e.
\begin{equation}
\mu_* = \mu - \Sigma^0. 
\end{equation}
As we present our results as functions of the net baryon density, the value
of $\Sigma^0$ (or the difference between $\mu$ and $\mu_*$) and its $\mu$
dependence need not be specified. More specifically, since we specify the 
total baryon density, we invert the expression of the density in terms of 
$\mu_*$ to obtain the baryon and anti-baryon Fermi-Dirac distributions (see
Eq.~(\ref{FD_dist}) below). In this way, the value of $\Sigma^0$ and its 
dependence on $\mu$ are not needed. With these approximations, the effective 
Hamiltonian of Eq.~(\ref{e:ham}) thus describes a system of quasi-particles 
with a momentum-dependent mass $M_*({\bf k})  = M - \Delta M({\bf k})$.

In the context of the model we just described, the free-space and in-medium 
creation and annihilation operators are related through a {\it fermionic} 
Bogoliubov-Valatin transformation~\cite{prc}
\begin{equation}
\left(
\begin{array}{c}
a_{\lambda,{\bf k}} \\
{\tilde a}^{\dagger}_{\lambda^\prime,-{\bf k}} \end{array}
\right) =
\left(
\begin{array}{cc}
c_{\bf k} &
\frac{f_{\bf k}}{|f_{\bf k}|} \,s_{\bf k} \,A  \\
-\frac{f^*_{\bf k}}{|f_{\bf k}|}\, s^*_{\bf k}\, A^{\dagger} &
c^*_{\bf k}
\end{array}
\right)
\left(
\begin{array}{c}
b_{\lambda,{\bf k}} \\
{\tilde b}^{\dagger}_{\lambda^\prime,-{\bf k}}
\end{array}      \right ),
\label{BVtransf}
\end{equation}
where $c_{\bf k}=\cos f_{\bf k}$, $s_{\bf k}=\sin f_{\bf k}$, $A$ is a 
$2~\times~2$ matrix with elements $A_{\lambda,\lambda^\prime}=
\chi^{\dagger}_{\lambda}\sigma\cdot{\hat{\bf k}}\tilde 
\chi^{\phantom\dagger}_{\lambda^\prime}$, where $\hat{\bf k} = 
{\bf k}/|{\bf k}|$, $\chi$ is a Pauli spinor and $\tilde \chi= -i\sigma^2 
\chi$. Here, $f_{\bf k}$ is the {\em squeezing} function. In the present case,
$f_{\bf k}$ is real and we will therefore drop the complex-conjugate notation 
in what follows. The effective Hamiltonian is brought to a diagonal form in 
the basis of the $b$ quanta when the squeezing function $f_{\bf k}$ is related
to the mass shift $\Delta M$ as~\cite{tfd},
\be
\tan(2 f_{\bf k}) = - \frac{|{\bf k}| \Delta M({\bf k}) }
{\omega^2_{\bf k} - M \Delta M({\bf k})}. 
\label{fk}
\ee  
With this, the thermal averages are then evaluated with the following density 
matrix operator
\begin{equation}
\hat\rho=\frac 1Z \exp\bigg(-\frac 1T \frac {V}{(2\pi)^3}\int ~d{\bf k}
\Bigl[(\Omega_{\bf k}-\mu_*)~b_{\lambda {\bf k}}^\dagger b_{\lambda,{\bf k}}+
(\Omega_{\bf k}+\mu_*)~{\tilde b}_{\lambda {\bf k}}^\dagger 
{\tilde b}_{\lambda,{\bf k}}\Bigr]
\bigg) \label{rho}.
\end{equation}
where $V$ is the volume of the system, $T$ the temperature, and 
$\Omega_{\bf k}=({\bf k}^2+{M_*}^2)^{1/2}$ is the quasi-particle energy. 
$Z$ is the trace of the exponential factor in Eq.~(\ref{rho}). 

The evaluation of the thermal averages in Eqs.~(\ref{gc}) and (\ref{gs}) is 
very easy and the results are 
\begin{eqnarray}
\langle a_{\lambda,{\bf k}}^\dagger\; a_{\lambda^\prime,{\bf k}^\prime}
\rangle &=&(c_{\bf k}^2 \;n_{\bf k}
+s_{\bf k}^2 \;(1-{\tilde n}_{\bf k}))\;\delta_{\lambda\lambda^\prime}
\;\delta_{{\bf k}{\bf k}^\prime} \label{ada},  \\
\langle \tilde a_{\lambda,-{\bf k}} \;a_{\lambda^\prime,{\bf k}^\prime},
\rangle &=& (1-{\tilde n}_{\bf k}-n_{\bf k})c_{\bf k}\;s_{\bf k}\; 
(A)_{\lambda\lambda^\prime}\;
\;\delta_{{\bf k}{\bf k}^\prime},
\label{aa}
\end{eqnarray}
where
\be
n_{\bf k}= \frac{1}{\exp\left[(\Omega_{\bf k} - \mu_*)/T\right] +1}~,~~~~~
{\tilde n}_{\bf k}= \frac{1}{\exp\left[(\Omega_{\bf k}+\mu_*)/T\right] +1}.
\label{FD_dist}
\ee 
The terms $\langle \tilde a^\dagger_{\lambda,{\bf k}} \; 
a_{\lambda^\prime,{\bf k}^\prime}\rangle$ and 
$\langle a^\dagger_{\lambda,{\bf k}} \;
\tilde a_{\lambda^\prime,{\bf k}^\prime}\rangle$ in Eq.~(\ref{rand}) vanish 
because they involve expectation values of two baryon annihilation
operators, or two anti-baryon creation operators, or one baryon and one 
anti-baryon operators.

\section{Spectra and correlations for mass-shifted fermions}

If a thermal gas of $b$ fermions freezes out {suddenly} at temperature $T$,
the observed single particle and anti-particle spectrum are given by
\begin{mathletters}
\begin{equation}
N_1({\bf k})  =  \frac{V}{(2 \pi)^3}\, \omega_{\bf k} \, \bigg[
c_{\bf k}^2\,n_{\bf k} + s_{\bf k}^2 (1 - \tilde n_{\bf k} )
\bigg ], 
\label{n1}
\end{equation}
\begin{equation}
{\tilde N}_1({\bf k})  =  \frac{V}{(2 \pi)^3}\, \omega_{\bf k} \, \bigg[
s_{\bf k}^2 (1 - n_{\bf k})+ c_{\bf k}^2\,{\tilde n}_{\bf k} 
\bigg ]\,.
\label{antin1}
\end{equation}
\label{sing}
\end{mathletters}
Note that the single-particle (anti-particle) spectrum includes a 
{\em squeezed} 
contribution. While the thermal part of the spectrum falls off exponentially 
for large values of $|{\bf k}|$, the squeezed contribution falls 
off only as a power of $|{\bf k}|$, because of the term proportional to
$s_{\bf k}^2$ in Eq.~(\ref{sing}).

For such a homogeneous system, the net baryon density $\rho_B$, the chaotic, 
and the squeezed amplitudes can be written from Eqs.~(\ref{ada}) and 
(\ref{aa}). They are given by
\bea
\rho_B & = &
\frac{g}{V}\sum_{\bf k}~\bigl( n_{\bf k} - \tilde n_{\bf k}\bigr), \\
G_c({\bf k}_1,{\bf k}_2) & = & \frac{V}{(2\pi)^3}\, \omega_{{\bf k}_1} \, 
[c_{{\bf k}_1}^2  \, n_{{\bf k}_1} 
+ s_{\bf k_1}^2 \,  (1- {\tilde n}_{{\bf k}_1})]\,
\delta_{{\bf k}_1,{\bf k}_2} , \label{Gc} \\
{\tilde G}_c({\bf k}_1,{\bf k}_2) & = & \frac{V}{(2\pi)^3}\,\omega_{{\bf k}_1}
\,[s_{\bf k_1}^2 \,(1- n_{{\bf k}_1})+c_{{\bf k}_1}^2  \, 
{\tilde n}_{{\bf k}_1}] \, \delta_{{\bf k}_1,{\bf k}_2} , \label{antiGc} \\
G_s({\bf k}_1,{\bf k}_2) &=& \frac{V}{(2\pi)^3}\, \omega_{{\bf k}_1} \, 
[(1- n_{{\bf k}_1} - {\tilde n}_{{\bf k}_1})  c_{{\bf k}_1} 
s_{{\bf k}_1} \,A^\dagger ] \,\delta_{{\bf k}_1,-{\bf k}_2} \,. \label{Gs}
\eea
There are two kinds of fermionic two-particle correlation functions.
Similarly to the bosonic case~\cite{acg}, let us denote by 
$C_2^{(++)}({\bf k}_1,{\bf k}_2)$ the case when the two particles are 
identical fermions, and by $C_2^{(+-)}({\bf k}_1,{\bf k}_2)$ when particle 1 
is a fermion and 2 is an anti-fermion. The correlation functions 
$C^{(--)}_2({\bf k}_1,{\bf k}_2)$ and $C^{(-+)}_2({\bf k}_1,{\bf k}_2)$ 
can be obtained from the (++) and $(+-)$ correlations by a trivial exchange 
of particle and anti-particle labels. For an infinite, homogeneous 
thermalized medium, these correlation functions are non-trivial only for 
identical or back-to-back momenta, ${\bf k}_2 = \pm {\bf k}_1$. 
We find that $C_2^{(++)}({\bf k},{\bf k}) = C_2^{(--)}({\bf k},{\bf k}) = 0$, 
the canonical value of the Fermi-Dirac correlation function, 
reflecting the anti-correlation of identical fermions, due to the 
Pauli exclusion principle.

The BBC for fermion  -- anti-fermion pairs given by 
Eq. (\ref{e:hbt}), reads as
\begin{eqnarray}
C_2^{(+-)}({\bf k},-{\bf k}) &=& 
1+ \frac{|G_s({\bf k},-{\bf k})|^2} {G_c({\bf k},{\bf k})
{\tilde G}_c(-{\bf k},-{\bf k})}
\nonumber \\
&=& 1 +\frac{(1 - n_{\bf k} - {\tilde n}_{\bf k})^2 (c_{\bf k}~s_{\bf k})^2}
{ \left[ c_{\bf k}^2 n_{\bf k} + s_{\bf k}^2 
(1-{\tilde n}_{\bf k}) \right]
\left[ c_{\bf k}^2 {\tilde n}_{\bf k} + s_{\bf k}^2 (1- n_{\bf k})\right]}. 
\label{BBCFa}
\end{eqnarray}
From this equation it follows that the fermionic BBC are 
unlimited from above. For sufficiently large values of $|{\bf k}|$, the 
Fermi-Dirac distribution $n_{\bf k}$ falls exponentially, 
while $s_{\bf k}$
decreases only as a power-law. Hence, for sufficiently large values of 
$|{\bf k}|$, the fBBC diverge as in the case of bosonic BBC, as 
$C^{(+-)}_2({\bf k}, - {\bf k}) \propto 1 + 1/s^2_{{\bf k}} \rightarrow 
\infty$. This divergence happens for small values of the mass-shift and for 
large values of ${\bf k}$ in both the fermionic and the bosonic cases. 

We next discuss numerical results. We use momentum-independent in-medium 
masses, as one would obtain in a mean-field approximation in model calculations
of the sort discussed in Refs.~\cite{wzm,franc}. There is no intrinsic 
difficulty in using momentum-dependent self-energies but a commitment to a 
specific model is then necessary. Since in this initial study we are more 
interested in the qualitative effects, rather than in precise, quantitative 
predictions, we simply use typical values for $M_*$ as obtained in mean-field 
calculations. In Fig.~1 we show the fBBC for $p \bar{p}$ pairs as a function 
of the in-medium mass $M_*=M-\Delta M$, for three illustrative values of the 
{\em net} baryonic density~$\rho_B$: the normal nuclear matter density, 
one tenth of the nuclear matter density, and $\rho_B=0$. This last value 
of $\rho_B$ corresponds to a baryon-free region, as expected to be formed 
at RHIC. We also show in this figure the corresponding result for
bBBC of $\phi$ mesons. Following Ref.~\cite{acg}, a finite time suppression 
factor is used to model a more gradual freeze-out. A detailed derivation for 
a gradual exponential freeze-out leads to a factor of the form~\cite{acg},
\begin{equation}
|\tilde F({\bf k})|^2 = \frac{1}{ [1+(2\,\Delta t\,\omega_{\bf k})^2]  },
\label{suppr}
\end{equation}
which multiplies $C_2^{(+-)}(k,k) -1$, the non-trivial part of the back-to-back
correlation function. We use $\Delta t = 2$~fm/s, as in
Ref.~\cite{acg}. We observe that the fBBC are strongly enhanced 
when the net baryonic density decreases, and that the shape of fBBC for 
$\rho_B=0$ is rather similar to bBBC. For a fixed temperature, as we increase 
the net baryon density obviously we have an excess of baryon over anti-baryons.
On the other hand, the fBBC will be larger for approximately equal baryon and 
anti-baryon densities and therefore the effect will be enhanced as we approach 
zero net baryon density. Our result shows that for a baryon free region, the 
BBCs are approximately independent of the bosonic or the fermionic nature of 
the particles. In addition, the fBBC are not only positive, as in the bosonic 
case, but they are also of the same order of magnitude as the corresponding 
bBBC.  

The expected magnitude of the effect is illustrated in Fig.~2 
for two, typical momenta of thermal-looking proton spectrum 
in Pb + Pb collisions at CERN SPS~\cite{na49-p}, for $|{\bf k}| = 500 $ MeV 
and for $|{\bf k}| = 800 $ MeV. As in Fig.~1, the finite time suppression 
factor of Eq.~(\ref{suppr}) is used. Fig.~2 indicates that the fBBC is 
strongly enhanced for increasing momentum $|{\bf k}|$. Fig.~2 also highlights 
that the magnitude of the fBBC is greatly enhanced as the net baryon density 
decreases from normal nuclear density to a vanishing value. Similarly to the 
bBBC case, the strength of fBBC is very sensitive to the shape of the 
freeze-out distribution, as well as to the value of the parameter 
$\Delta t$, the particle freeze-out duration, as discussed in Ref.~\cite{acg}.
Note that in the limit of infinitely slow freeze-out, $\Delta t \rightarrow
\infty$, both fBBC and bBBC vanish. 

It is particularly interesting to compare the BBC discussed above in 
Eq.~(\ref{BBCFa}) with Eq.~(19) of Ref.~\cite{acg} for bosonic case as
\begin{eqnarray}
C_2^{(+-)}({\bf k},-{\bf k}) & = &  1 + \frac{(1+ n^b_{{\bf k}} + 
n^b_{{\bf k}})^2 \,
(c^b_{{\bf k}}\, s^b_{-{\bf k}})^2}{ \left[ (c^b_{{\bf k}})^2 \,
n^b_{{\bf k}} + (s^b_{-{\bf k}})^2 \,(1+ n^b_{{\bf k}}) \right]
\left[(c^b_{{\bf k}})^2 \,n^b_{{\bf k}} + (s^b_{{\bf k}})^2\,  
(1 + n^b_{{\bf k}}) \right] }\,,
\label{BBCFB} \\
n^b_{{\bf k}} & = & \frac{1}{\exp{(\Omega^b_{{\bf k}} /T)}  - 1} \,.
\end{eqnarray}
Note, however, that the symbols $s_{\bf k}$ and $c_{\bf k}$
are related to $\sin f_{\bf k} $ and $\cos f_{\bf k} $, as defined by 
Eq.(\ref{fk}) in the fermionic case, while in the bosonic case, these symbols
mean $s^b_{{\bf k}}=\sinh f^b_{\bf k}$ and $c^b_{{\bf k}B}=\cosh f^b_{\bf k}$,
where $f^b_{\bf k} = \frac{1}{2} \log{(\omega_{\bf k}/\Omega_{\bf k})}$, as 
given by Eq.~(11) of Ref.~\cite{acg}, is also real. It follows that the BBC 
diverge with increasing $|{\bf k}|$ as the inverse of the single-particle 
spectra, both  in the fermionic and bosonic cases.

\section{Conclusion} BBC stands for  Back-to-Back Correlations of particle and 
anti-particle pairs. These correlations appear if in-medium interactions 
lead to the modification of hadronic masses in the medium.  The origin of 
these back-to-back correlations is an entirely quantum effect, related to 
the propagation of particle fields through a space-like boundary surface
between the medium and the asymptotic region. The strength of these 
correlations can be unlimitedly large, and the shape of the BBC is similar 
for fermions and bosons. Deep mathematical and physical reasons 
are behind these similarities. A sudden freeze-out of thermalized 
medium-modified quanta to asymptotic fields is described by a bosonic or 
a fermionic Bogoliubov transformation. Although these transformations are 
canonical, they connect Fock spaces that become unitarily inequivalent in 
the infinite volume limit and the state of the medium corresponds to strongly 
correlated, squeezed particle--anti-particle states of the asymptotic quanta. 

At large values of momenta, the similarity of fermionic BBC to bosonic 
BBC reflects a symmetry of the decaying medium to observable 
boson - anti-boson and fermion - anti-fermion pairs. This effect vanishes 
in case of exactly zero in-medium mass-modification, but for large values 
of $|{\bf k}|$ and small values of mass-shifts, BBC can be very large and
should be observable if the freeze-out process of the medium modified
quanta is sufficiently sudden, for example if the duration of particle
freeze-out is within $2$~fm/c. 

In this Letter, we have extended the concept of back-to-back correlations 
to an important new domain of broad experimental and theoretical interest.
As a reduction of the net baryon density by a factor of 10 increases 
dramatically the magnitude of the effect, the almost baryon-free mid-rapidity 
region of $Au + Au$ collisions at  RHIC seems to be an ideal place to find 
the back-to-back correlations of proton - anti-proton or 
$\Lambda$ - $\overline{\Lambda}$ pairs experimentally. 

\acknowledgments
We thank M. Asakawa, M. Gyulassy and G. Zim\'anyi for useful comments and 
suggestions during the completion of the manuscript.  T. Cs\"org\H o is 
grateful to Y. Hama and S. S. Padula for creating an inspiring working 
atmosphere during his visit in S\~ao Paulo and to M.  Gyulassy for the kind 
hospitality at the Columbia University. This research has been supported
in part by CNPq and FAPESP grants 98/2249-4, 99/08544-0, 99/09113-3, by the 
Hungarian OTKA T025435, T029158, the US - Hungarian Joint Fund MAKA 652/1998
and the NWO-OTKA grant N025186, by the US Department of Energy grants 
DE - FG02 - 93ER40764, DE - FG02 - 92 - ER40699, DE - AC02 - 76 - CH00016 
and by a Bolyai Fellowship.
%

%
\vfill\eject
\vspace{-1.5cm}
\begin{figure}
\begin{center}
\hspace{0.2cm}
\epsfxsize=14.5cm\epsfbox{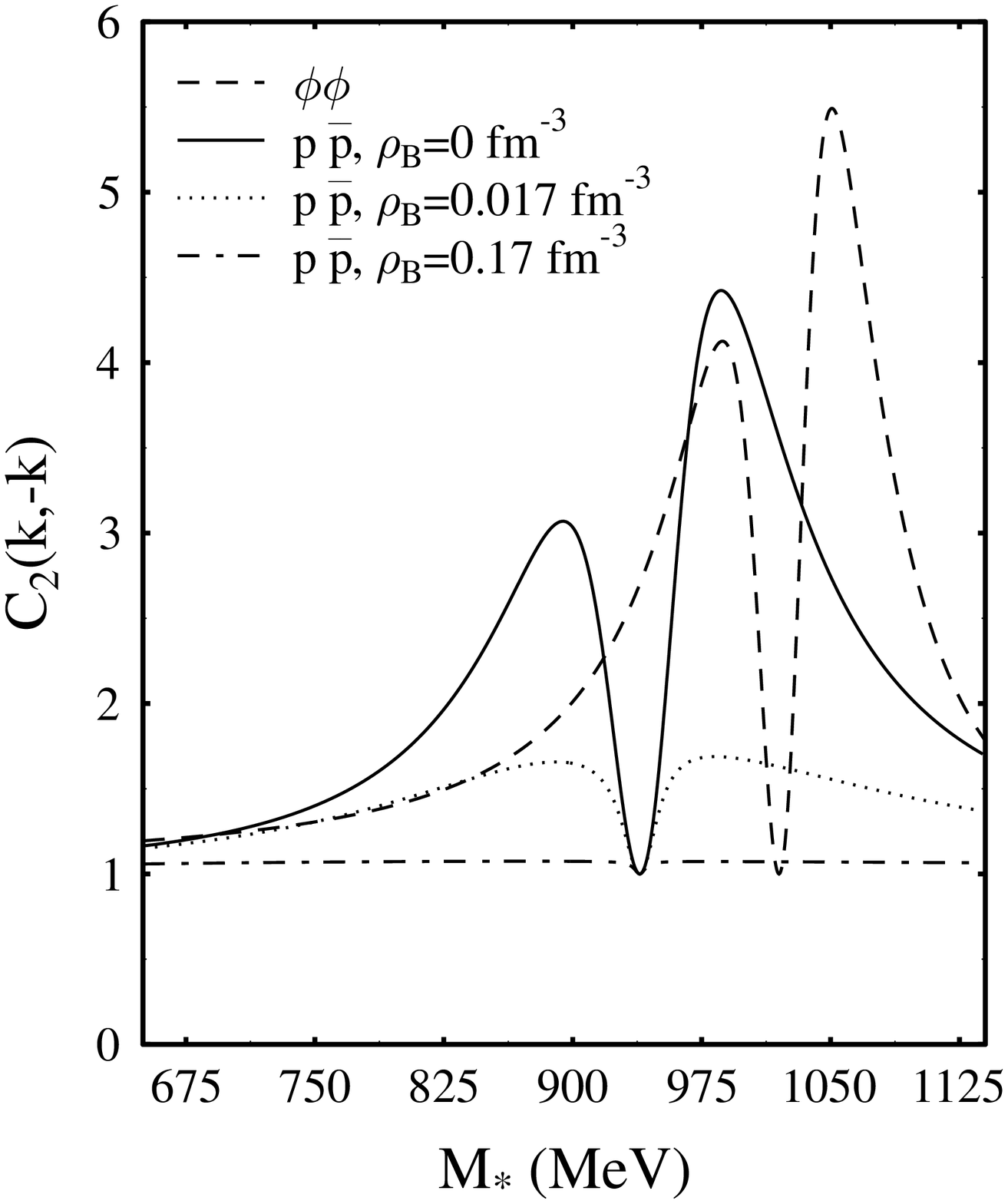}
\end{center}
\caption{Back-to-back correlations of proton - anti-proton pairs
and $\phi$-meson pairs, for $T=140$~MeV, $\Delta t =2 $ fm/c and
$|{\bf k}| = 800$ MeV/c.}
\label{f:bbcf-1}
\end{figure}
\vfill\eject
\begin{figure}
\begin{center}
\hspace{0.2cm}
\epsfxsize=14.5cm\epsfbox{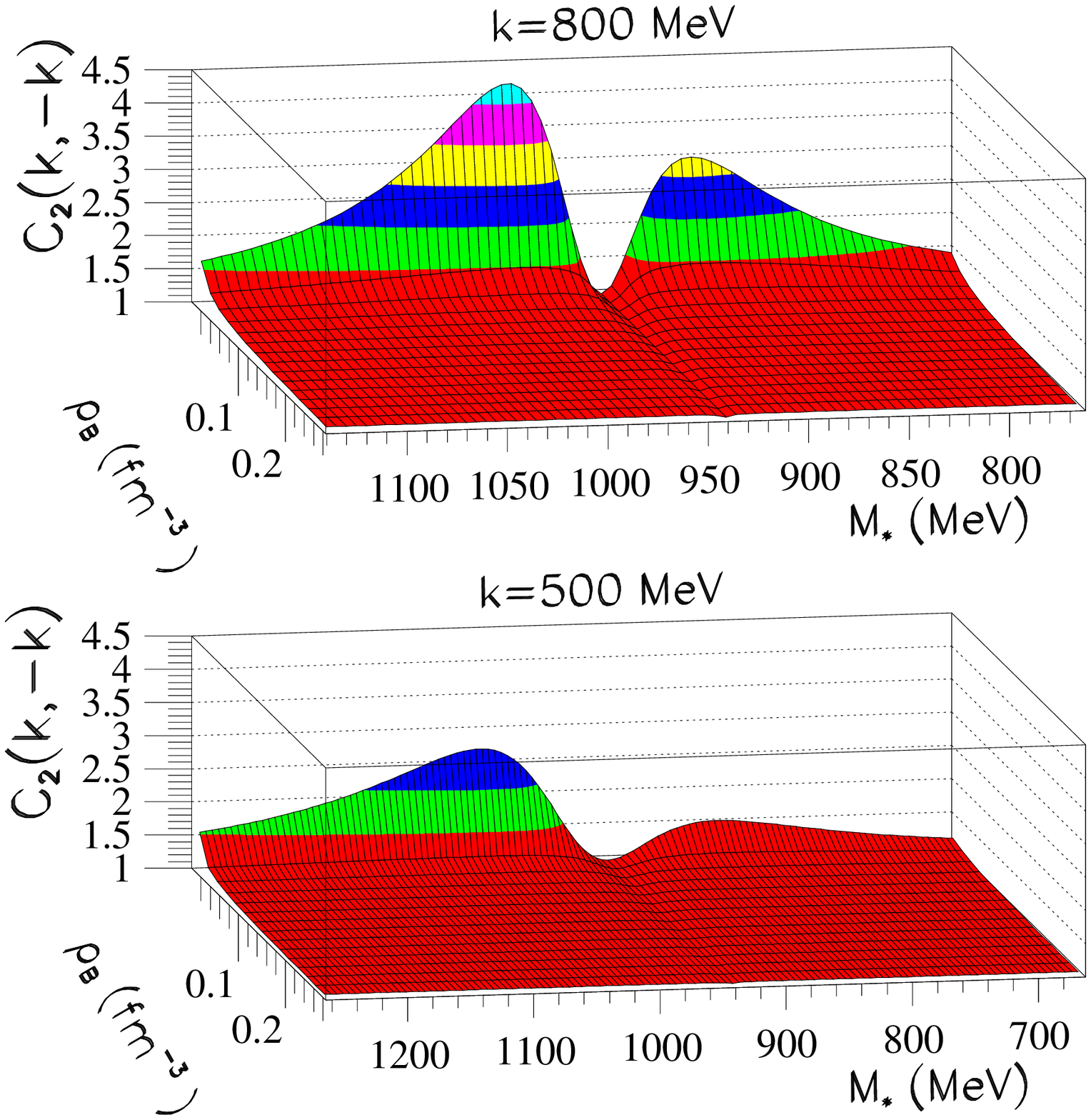}
\end{center}
\caption{Dependence of the fBBC on the in-medium modified proton mass, $M_*$, 
and on the net baryon density, $\rho_B$, for $T=140$~MeV, 
$\Delta t$ = 2 fm/c 
and two typical values of the momentum $|{\bf k}|$. 
The fBBC strongly increases 
with decreasing net baryon density and increasing values of $|{\bf k}|$.}
\label{f:bbcf-2}
\end{figure}

\begin{thebibliography}{99}
\bibitem{ac} M. Asakawa and T. Cs\"org\H o, hep-ph/9612331, Heavy Ion 
Physics {\bf 4}, (1996) 233; M. Asakawa and T. Cs\"org\H o, quant-ph/9708006, 
in proc. Strong and Electroweak Matter'97 Conference, Eger, Hungary, 
May  1997. (World Scientific, Singapore, 1998,  F. Csikor  and Z. Fodor eds.) 
p. 332.
%
\bibitem{acg} M. Asakawa, T. Cs\"org\H o and M. Gyulassy, Phys. Rev. Lett. 
{\bf 83}, (1999) 4013.
%
\bibitem{andreev} A. F. Andreev. Sov. Phys. JETP {\bf 19}, (1964) 1228;
see e.g. H. Hoppe, U. Z\"ulicke and G. Sch\"on, Phys. Rev. Lett. {\bf 84}, 
(2000) 1804; J. Torr\`es and T. Martin, Eur. J. Phys. {\bf B12} (1999) 319.
\bibitem{bl-qm99} A. Ster, T. Cs\"org\H{o} and B. L\"orstad, Nucl. Phys. 
{\bf A661}, (1999) 419c.
%
\bibitem{wzm} J.D. Walecka, Ann. Phys. (N.Y.) {\bf 83}, (1974) 491;
J.~Zim\'anyi and S. Moszkowski, Phys. Rev. C. {\bf 42}, (1990) 1416.
%
\bibitem{franc} A. Bouyssy, J.-F. Mathiot, N.V. Giai and S. Marcos
Phys. Rev. C {\bf 36}, (1987) 380.
%
\bibitem{prc} A. Mishra, P.K. Panda, S. Schramm, J. Reinhardt, W. Greiner,
Phys. Rev. C {\bf 56}, (1997) 1380.
%
\bibitem{tfd}A.L. Fetter J.D. Walecka, {\em Quantum Theory of 
Many-Particle Systems} (McGraw-Hill, NewYork, 1971).
%
\bibitem{na49-p} H. Appelshauser et al, NA49 Collaboration, Phys. Rev. Lett. 
{\bf 82}, (1999) 2471.
\end{thebibliography}
\end{document}